\documentclass{iopart}
\usepackage{graphicx}

\begin{document}
\title[Global-Local Quantization]{Pointwise reconstruction of wave functions from their moments through weighted polynomial expansions: an alternative global-local quantization procedure}
\author{Carlos  R. Handy$^1$, Daniel Vrinceanu$^1$, Carl Marth$^2$, and Harold A. Brooks$^1$}
\address{Department of Physics, Texas Southern University, Houston, Texas 77004; \\$^2$ Dulles High School,Sugar Land, Texas 77459}
\ead{handycr@tsu.edu}

\begin{abstract}
Many quantum systems admit an explicit analytic Fourier space expansion, besides the usual  analytic Schrodinger configuration space representation. We argue that the use of weighted orthonormal polynomial expansions for the physical states (generated through the power moments)  can define  an $L^2$ convergent, non-orthonormal, basis expansion with sufficient point-wise convergent behaviors enabling the direct coupling of the global (power moments) and local (Taylor series) expansions in configuration space. Our formulation is elaborated within the orthogonal polynomial projection quantization (OPPQ) configuration space representation previously developed ({\em J. Phys. A: Math. Theor.} {\bf{46}} 135202). The quantization approach pursued here defines an  alternative strategy emphasizing the relevance OPPQ  to the reconstruction of the local structure of the physical states.
\end{abstract}
\submitto{Physics Letters} 
\pacs{03.65.Ge, 02.30.Hq, 03.65.Fd}

\vfil\break
\section{Introduction}

Many quantum systems are defined by analytic solutions which in turn compell us to generate them through (approximate) analytic methods. Although one is generally more interested in the hierarchical  large scale structure leading to quantization (for the bound states), recovering the local structure is also important and the principal focus of this work.  Our methods have the important advantage that they are essentially algebraic in nature, allowing for high precision calculations through multiple precision algorithms such as Mathematica. 

 Of interest to us are those systems for which the power series expansion in momentum space, and the local power series expansion in configuration space, are known in the sense that their recursive structure can be generated from the very nature of the underlying Schrodinger equation. Our objective is to define a robust quantization procedure that directly couples  both sets of power series coefficients. 

Since the momentum space expansion is governed by the power moments of the solution, $\mu_p = \int dx x^p\Psi(x)$, our approach is tantamount to defining an effective function-moment reconstruction ansatz, which is generally a difficult problem. One advantage we have is that we know the algebraic structure for all the power moments, and the underlying solutions are generally smooth and bounded, as opposed to the more general problem.  In keeping with this, mathematicians have known that the use of weighted (orthonormal) polynomial expansions,
\begin{eqnarray}
\Psi(x) = \sum_n \Omega_n P_n(x) {\cal R}(x), \nonumber
\end{eqnarray}
 particularly for Freudian weights, ${\cal R}(x) = e^{-|x|^q}$, $q > 1$, can converge (pointwise) to the types of solutions encountered for physical systems [1]. The expansion coefficients, $\Omega_n$, are determined by the power moments, as discussed  below. We stress that the expressions $\{P_n(x) {\cal R}(x)\}$ represent a non-orthonormal basis, which is an essential component to the flexibility of the above representation as used here.

Despite the extensive mathematical literature on the importance of weighted polynomial expansions, its relevance for quantizing physical systems has not been appreciated. We believe that this stems from the preferance physicists have for configuration space based bound state quantization analysis, as opposed to those based in Fourier space and (when appropriate) dependent on the underlying power moment structure.  

Recently, the Orthogonal Polynomial Projection Quantization (OPPQ) method was developed [2], using weighted polynomial expansions for quantization, and emphasizing a numerical approximation to the asymptotic condition $\lim_{n\rightarrow \infty}\Omega_n = 0$, which is a defining condition for the bound states. A particularly important achievement of the OPPQ analysis is that it is exceptionally stable and rapidly converges for weights (or {\it{reference}} functions) inaccesible to other methods, such as the Hill determinant method [3]. As  argued by Hautot [4], and confirmed by Tater and Turbiner (for the sextic anahrmonic potential) [5] if one uses reference functions that mimic the asymptotic form of the physical states, the Hill determinant method can become unstable, nonconvergent, or converge to the wrong solution. The OPPQ approach is not plagued by these problems.  Since the Tater and Turbiner analysis does not clearly show these behaviors, we reproduce them here, as given in Fig 1 (Hill determinant results). The relevant discussion is given below, as well as the comparative power of OPPQ and the alternate``global-local" quantization method presented here (see Table I), for these anomalous Hill determinant cases of non-convergence.

The OPPQ approach is solely dependent on the power moments. The mathematical theorems strongly support, for appropriate weights, the pointwise convergent properties of such representations. Given this, a natural question is can we quantize by directly coupling the OPPQ, moment based representation, to the local structure of the wavefunction, as determined by its Taylor series expansion?  In this work, we examine the effectiveness of such a ``global-local" quantization approach.  The stability of such a procedure is further confirmation of the exceptional numerical stability properties of the OPPQ representation.  

There is another related  result derived from the scaling transform formalism underlying wavelet analysis (${\cal N}_a \int dx {\cal S}({x\over a})  = 1$) [6]. It also couples the power moments to the local structure of the wavefunction. Consider the scaling transform of the wavefunction, for some apppropriate, bounded, scaling function, ${\cal S}(x) \equiv \sum_n \sigma_n x^n$:

\begin{eqnarray}
{\underline{S\Psi}}(a,b) = {\cal N}_a \int dx  \ {\cal S}({{x-b}\over a}) \Psi(x).
\end{eqnarray}
Depending on the asymptotic decay of the scaling function relative to the physical solution, the scaling transform will be analytic in the inverse scale variable, as generated from the power moments:
\begin{eqnarray}
{\underline{S\Psi}}(a,b) = {\cal N}_a \sum_n {{\sigma_n}\over {a^n}} \mu_n(b),
\end{eqnarray}
where $\mu_n(b) = \int dx x^n \Psi(x+b)$, involving a linear combination of the power moments for $b = 0$.  Thus, the power moments control the large scale structure of the scaling transform. Recovery of the local properties then requires the small scale asymptotic analysis

\begin{eqnarray}
{\underline{S\Psi}}(a,b) & = &  a{\cal N}_a \int dx  {\cal S}(x) \Psi(ax +b)  \nonumber \\
\lim_{a\rightarrow 0} {\underline{S\Psi}}(a,b) & \approx &  {1\over{\nu_0}} \sum_n {a^n }{{\Psi^{(n)}(b) \nu_n}\over{n!}},
\end{eqnarray}
where $\nu_n = \int dx x^n{\cal S}(x)$.  

For all one dimensional systems of the type considered here, upon solving for the physical moments of the bound state solutions, one can analytically continue the scaling transform and recover excellent pointwise results for the wavefunction [6].  That is, knowledge of the physical power moments (derived by other means) coupled with the appropriate analytic continuation strategy, proved very effective in recovering the local solution. However, imposing the local structure (at the turning points) on the scaling transform representation, in order to determine the physical power moments from the local derivatives, proved ineffective. As demonstrated here, the OPPQ representation does allow for this type of ``global-local" analysis.
\vfil\break

\section {OPPQ and Global-Local Quantization}

In order to make more precise the above claims, consider the one dimensional Fourier transform, assumed to be analytic (usually entire), with a corresponding $k$-space power series:

\begin{eqnarray}
{\hat \Psi}(k) & =  &{1\over {\sqrt{2\pi}}}\int dx \ e^{-ikx} \Psi(x) \\
 & = & {1\over {\sqrt{2\pi}}}\sum_{p=0}^\infty {{\mu_p}\over {p!}} (-ik)^p.
\end{eqnarray}
  We limit our analysis to (multidimensional) quantum systems for which the  moments can be generated through a linear recursive relation of order $1+m_s$ (in the one dimensional case), referred to as the {\it{moment equation}}.  In many cases, a coordinate transformation may be necessary to realize this. Its  structure will take on the form

\begin{eqnarray}
\mu_p = \sum_{\ell = 0}^{m_s} M_{p,\ell}(E) \mu_\ell,
\end{eqnarray}
where $m_s$ is problem dependent. The $M_{p,\ell}(E)$'s  are known functions of the energy. The unconstrained moments $\{\mu_0,\ldots, \mu_{m_s}\}$ are referred to as the {\it {missing moments}}.

The configuration space wavefunction is to be represented as 
\begin{eqnarray}
\Psi(x) = \Big(\sum_{n=0}^\infty c_n x^n \Big) {\cal R}(x),
\end{eqnarray}
 for some positive weight function, ${\cal R}(x) > 0$. The nature of the potential, and domain, usually dictates the possible choices for the weight.  We are assuming that this is done in such manner ensuring that the series expression in Eq.(7) corresponds to an analytic expansion near the origin (or any other desired point).  Under these assumptions, the $c_n$ coefficients  satisfy a recursive, second order, Frobenius method relation
\begin{eqnarray}
c_n = T_{n,0}(E) c_0 +  T_{n,1}(E)  c_1.
\end{eqnarray}

Given that the global power moments, $\mu_p$, and the local $c_ n$, satisfy known, energy dependent constraints, respectively, how can we quantize by constraining both sets of variables?  

\subsection{OPPQ}

What is clearly needed is a robust, wavefunction representation ansatz capable of recovering (in a stable manner) the local structure of the wavefunction from the global, power moments. Weighted polynomial expansions,as implemented within the OPPQ representation, can provide this.  Specifically,  we will work with

\begin{eqnarray}
\Psi(x) = \sum_{j=0}^\infty \Omega_j P_j(x) {\cal R}(x),
\end{eqnarray}
where the orthonormal polynomials, $P_j(x) \equiv \sum_{i=0}^j \Xi_i^{(j)} x^i$,  satisfy 
\begin{eqnarray}
\langle P_{j_1} | {\cal R}|P_{j_2} \rangle = \delta_{j_1,j_2}.
\end{eqnarray}
The projection coefficients are easily generated :

\begin{eqnarray}
\Omega_j  & = & \int dx \ \Psi(x) P_j(x) , \\
& = & \sum_{i=0}^j \Xi_i^{(j)} \mu_i ,
\end{eqnarray}
or
\begin{eqnarray}
\Omega_j (\mu_0,\ldots,\mu_{m_s})  =  \sum_{\ell = 0}^{m_s} \Big( \sum_{i =0}^j   \Xi_i^{(j)} M_{i,\ell}(E) \Big) \mu_\ell.
\end{eqnarray}

If the weight ${\cal R}(x)$ satisfies the condition that  $\int dx\  {{{\Psi^2}(x)}\over{ {\cal R}(x)}} < \infty$, then one can argue that [2] 
\begin{eqnarray}
Lim_{j \rightarrow \infty} \Omega_j = 0.
\end{eqnarray}
This analysis is repeated, in a more complete manner, in Sec. IV.
These conditions hold, in particular, if the weight decays, asymptotically, no faster than the physical solutions:
$\lim_{|x|\rightarrow \infty}{ {|\Psi(x)|}\over {{\cal R}(x)}} < \infty$.

Within the original OPPQ quantization analysis, we approximate this asymptotic limit by taking $\Omega_{n} = 0$ for $ N-m_s \leq n \leq N$, $N \rightarrow \infty$. This leads  to an  $(m_s+1) \times (m_s +1)$  determinantal constraint on the energy, yielding rapidly converging approximations to the physical energies.

This approach allows for great flexibility in how the weight, or {\it {reference function}},  is chosen. In particular, one can allow the weight to mimic the asymptotic form of the physical solution, or take on the form of any positive solution (i.e. the bosonic ground state, if known, even approximately). This is in sharp contrast to the popular Hill determinant approach corresponding to taking  $c_{N} = 0, c_{N-1} = 0$ in Eq.(8), and letting $N \rightarrow \infty$.

Within the Hill determinant approach, truncating the $c$-power series is relevant if the system is exactly, or quasi-exactly, solvable. One might then believe that this works as an approximation in the general case. Initial studies of the Hill determinant approach confirmed this, for weights that did not mimic the physical asymptotic form of the solution.

One simple observation that suggests potential problems with the Hill determinant approach is that the asymptotic behavior of the $c$'s cannot be directly related to the normalizable, or non-normalizable, behavior of the physical or unphysical states, respectively. Indeed,  it was pointed out by Tater and Turbiner that the Hill determinant fails to converge, or converges to the wrong energy levels, in cases where the reference function is chosen to mimic the physical asymptotic form.  They used the sextic anharmonic potential as an example: $V(x) = ax^2+bx^4+x^6$. The asymptotic form of the wavefunction is ${\cal R}(x) = e^{-(x^4+bx^2)/4}$.  The corresponding recursion relation for the $c$'s is

\begin{eqnarray}
c_{n+2} = {{(a+2n-1-b^2/4) c_{n-2} + \left(b(n+1/2)-E\right) c_n}\over{(n+1)(n+2)}}. 
\end{eqnarray}

Consistent with the Tater and Turbiner results [5], Hautot [4] had argued that such finite difference relations, coupled with the Hill determinant conditions ($c_N = 0$, $N = even$, $N \rightarrow \infty$, for parity invariant systems) can fail to take into account certain dominant solutions essential to quantization. He proposed a complicated procedure for fixing this problem. We believe that OPPQ is a more transparent solution that achieves the same result.

\begin{figure}[h]
\includegraphics[width=5.in]{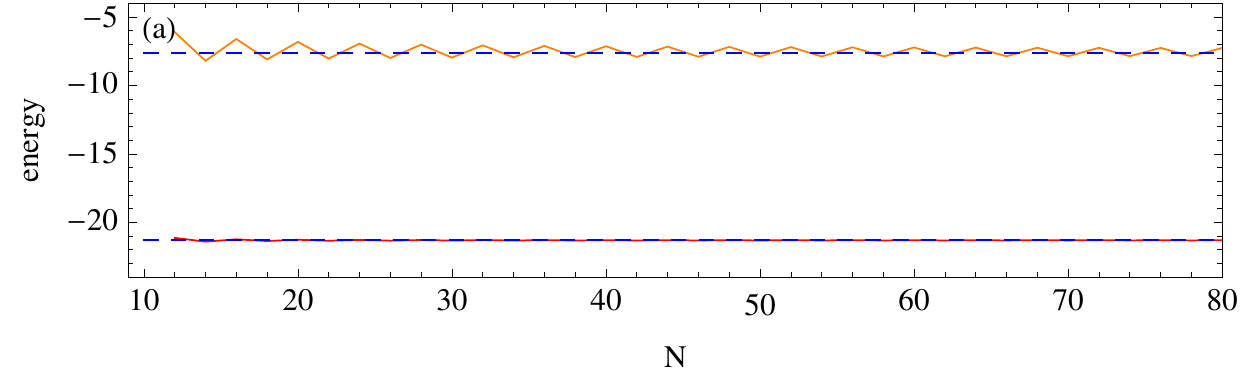}\\
\includegraphics[width=5.in]{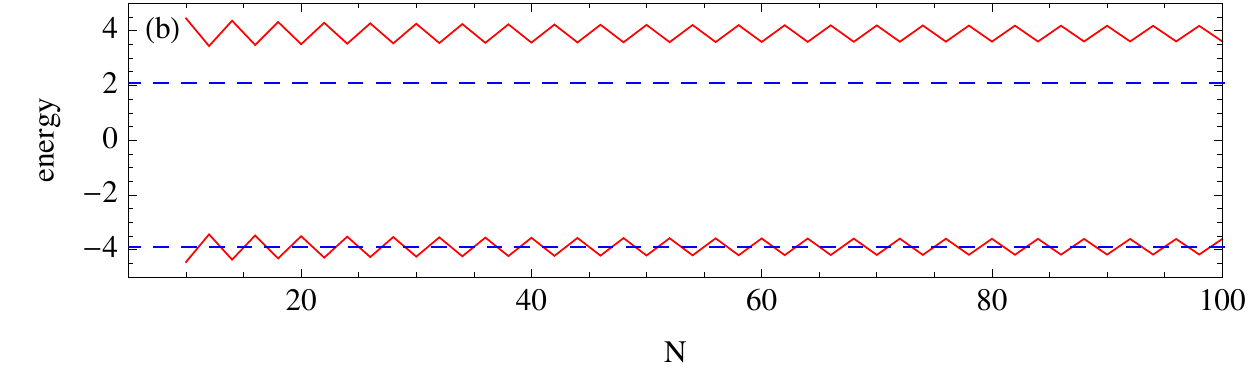}\\
\includegraphics[width=5.in]{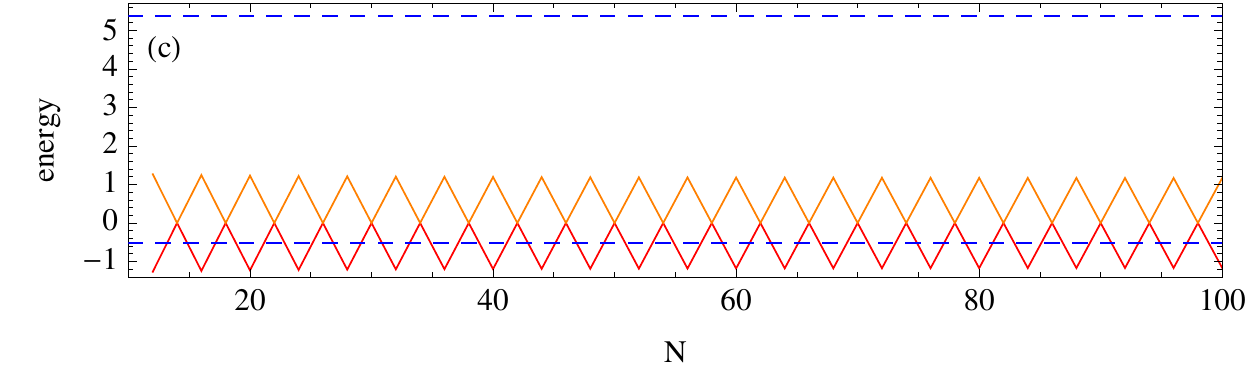}\\
\caption{\label{fig1}
Hill determinant results for anomalous cases of the sextic anharmonic oscillator for parameter values:
(a) (a=-18, b=0), (b) (a=-8, b=0) and (c) (a=-4, b=0).
}
\end{figure}

For illustrative purposes we note that in Fig \ref{fig1}(a-c)  we give the convergence of the first two even energy levels for three parameter cases:
($a=-18$, $b=0$), ($a=-8$, $b=0$) and ($a=-4$, $b=0$). In the first case both energy levels appear to 
converge to the correct limit, albeit at a very slow rate. For the second case only the ground state shows 
correct convergence, while the first excited state converges to the wrong limit. In the third situation the
Hill determinant method fails for all energy levels, there is no convergence and in some cases there are no
real solutions, represented by a value of ``0" in Figure. 1c. 
All three cases are correctly recovered by the global-local OPPQ variant developed in this paper (described in Sec. 2.2), as given in Table 1. The indicated limits are in keeping with a pure OPPQ analysis as publshed elsewhere [2].

\begin{table}
\caption{\label{table-convergence}
Convergence properties for $V(x) = a x^2 + b x^4 + x^6$ using $R(x) = e^{-x^4/4}$ by
using the global-local quantization.
}
\centering
\begin{tabular}{ccccc}
 $N$ & $E_0$ & $E_1$ & $E_2$ & $E_3$ \\
 \hline
 \multicolumn{5}{c}{$a = -18$, $b = 0$}\\ 
 \hline
 10 & -23.943914597 & -17.917277160 & -11.400078722 & -8.480104521 \\
 20 & -21.324952739 & -21.322438503 & -7.599903464 & -7.359503437 \\
 30 & -21.323394711 & -21.321841616 & -7.599035456 & -7.360657993 \\
 40 & -21.323394694 & -21.321841620 & -7.599035461 & -7.360657990 \\
 \hline
 \multicolumn{5}{c}{$a = -8$, $b = 0$}\\ 
 \hline
 10 & -5.477684656 & -3.531662882 & 1.811585056 & 6.570677286 \\
 20 & -3.900838586 & -3.534341976 & 2.086375864 & 6.054510025 \\
 30 & -3.900635158 & -3.534354171 & 2.086528016 & 6.055405087 \\
 40 & -3.900635159 & -3.534354170 & 2.086528012 & 6.055405205 \\
 \hline
 \multicolumn{5}{c}{$a = -4$, $b = 0$}\\ 
 \hline
 10 & -0.625342803 & 1.038625070 & 5.801273017 & 8.725547624 \\
 20 & -0.523263742 & 1.005832318 & 5.374951631 & 10.5614373902 \\
 30 & -0.523268623 & 1.005768335 & 5.374969926 & 10.572585458 \\
 40 & -0.523268622 & 1.005768340 & 5.374970009 & 10.572585045 \\
\end{tabular}
\end{table}

In  general, within the OPPQ ansatz,  the better the reference function mimics the asymptotic form of the solution, the faster the convergence to the physical energies.  Other redeeming features of the OPPQ procedure are that  the reference function need not be analytic. Thus, for the quartic anharmonic potential, $V(x) = x^4 + mx^2$, converging results are obtained if ${\cal R}(x) = e^{-{{|x|^3}\over 3}}$. In principle, for this case, a differentiable form of the reference function would be ${\cal R}(x) = {1\over {exp({{{x^3}\over 3} })+ exp({-{{x^3}\over 3}}})}$. Also, for generating the energies, one does not need the explicit, configuration space, representation for the reference function. Thus one could use the (unknown)  bosonic ground state (which must be positive), provided  its power moments can be generated to high accuracy, enabling the generation of its corresponding orthonormal polynomials. In principle, the Eigenvalue Moment Method could be used for such cases (generating high precision values for the power moments of the ground state, as well as the energy, through converging bounds) [7-9]. This convex optimization procedure defines the first [10] application of  semidefinite programming analysis to quantum operators [7], and its computational implementation through linear programming [8,9,11].

We note that a more conventional analysis involving expanding $\Psi(x)$ in terms of an orthonormal basis ${\cal P}_n(x) {\cal R}^{1\over 2}(x)$, or 
$\Psi(x) = \sum_n \gamma_n {\cal P}_n(x) {\cal R}^{1\over 2}(x)$ , does not provide the flexibility of OPPQ, since the generation of the projection coefficients involves the integrals $\gamma_n =\int dx \  {\cal P}_n(x){\cal R}^{1\over 2}(x)\Psi(x)$, which cannot be expressed, generally, as a known (i.e. in  closed form)  linear combination of the power moments of $ {\cal R}^{1\over 2}(x)\Psi(x)$, except for special weights. One good example is the aforementioned observation that OPPQ allows the use of the (accurately determined) power moments of the (bosonic) ground state, for quantizing the excited states. This type of analysis is not possible within the more conventional, ortho-normal basis, approach.

A final observation is that in selecting the appropriate ${\cal R}(x)$ that duplicates the  asymptotic behavior of the configuration space  solution, we expect its Fourier transform to also  mimic the asymptotic decay of the Fourier transform of the physical solution. This is generally consistent with the uncertainty principle relation.

\subsection{Global-Local Quantization}

The OPPQ expansion in Eq.(9) was originally developed  in the spirit of a (non-orthonormal) basis expansion where the expansion coeffcients are given by Eq.(11)  and the following integral expression is finite:  $\int dx \ {{{\Psi^2}(x)}\over{{\cal R}(x)}} = \sum_j \Omega_j^2 < \infty$. There was no demand for pointwise convergence. However, there are good mathematical reasons for expecting the OPPQ representation to be (non-uniformly) convergent in a pointwise manner. As previoulsy noted, this representation is a specific case of the more general problem of representing analytic functions by weighted families of polynomials. For Freudian weights of the form ${\cal R}(x) = e^{-|x|^q}$, $q > 1$, it is known [1] that the representation in Eq.(9) converges within an infinite strip in the complex-$x$ plane whose width is determined by the closest singularity (of the physical solution) to the real axis. If we assume this, in general, for the types of physical systems of interest, then the natural question is to test the pointwise convergence of such representations at the origin (among other possibilities):

\begin{eqnarray}
{{\Psi(x)}\over {{\cal R}(x)}} = \sum_{j=0}^\infty \Omega_j P_j(x) = \Big(\sum_{n=0}^\infty c_n x^n\Big) , 
\end{eqnarray}
where
\begin{eqnarray}
\sum_ {j = n}^\infty \Xi_n^{(j)} \Omega_j  = c_n = T_{n,0}(E) c_0+ T_{n,1}(E) c_1. 
\end{eqnarray}
We approximate this through the truncation

\begin{eqnarray}
\sum_ {j = n}^N  \Xi_n^{(j)}\Omega_j(\mu_0,\ldots,\mu_{m_s})\  = c_n = T_{n,0}(E) c_0+ T_{n,1}(E) c_1,
\end{eqnarray}
where we make explicit $\Omega$'s linear dependence on the missing moments. Since there are $m_s+3$ linear variables $\{c_0,c_1,\mu_0,\ldots,\mu_{m_s}\}$, we must take $n =0, \ldots, m_s+2$, in Eq.(18), although $N \rightarrow \infty$. An energy dependent, $(m_s+3) \times (m_s+3)$ determinantal equation ensues, yielding converging results for the physical energies, as $N \rightarrow \infty$.

The above analysis was implemented on the anomalous parameter values for the sextic anharmonic oscillator, as calculated through the Hill determinant approach given in Fig. 1. Table I shows the exceptional stability of the above ``global-local" quantization procedure. These results agree with a pure OPPQ analysis as given in reference [2]; thereby strongly affirming the reliability of the OPPQ representation in capturing the local behavior of the physical solutions. A second example is given below.

\vfil\break

\section{Global-Local Quantization for a Rational Anaharmonic Oscillator} 

\begin{figure}[h]
\centering
\includegraphics[width=3.0in]{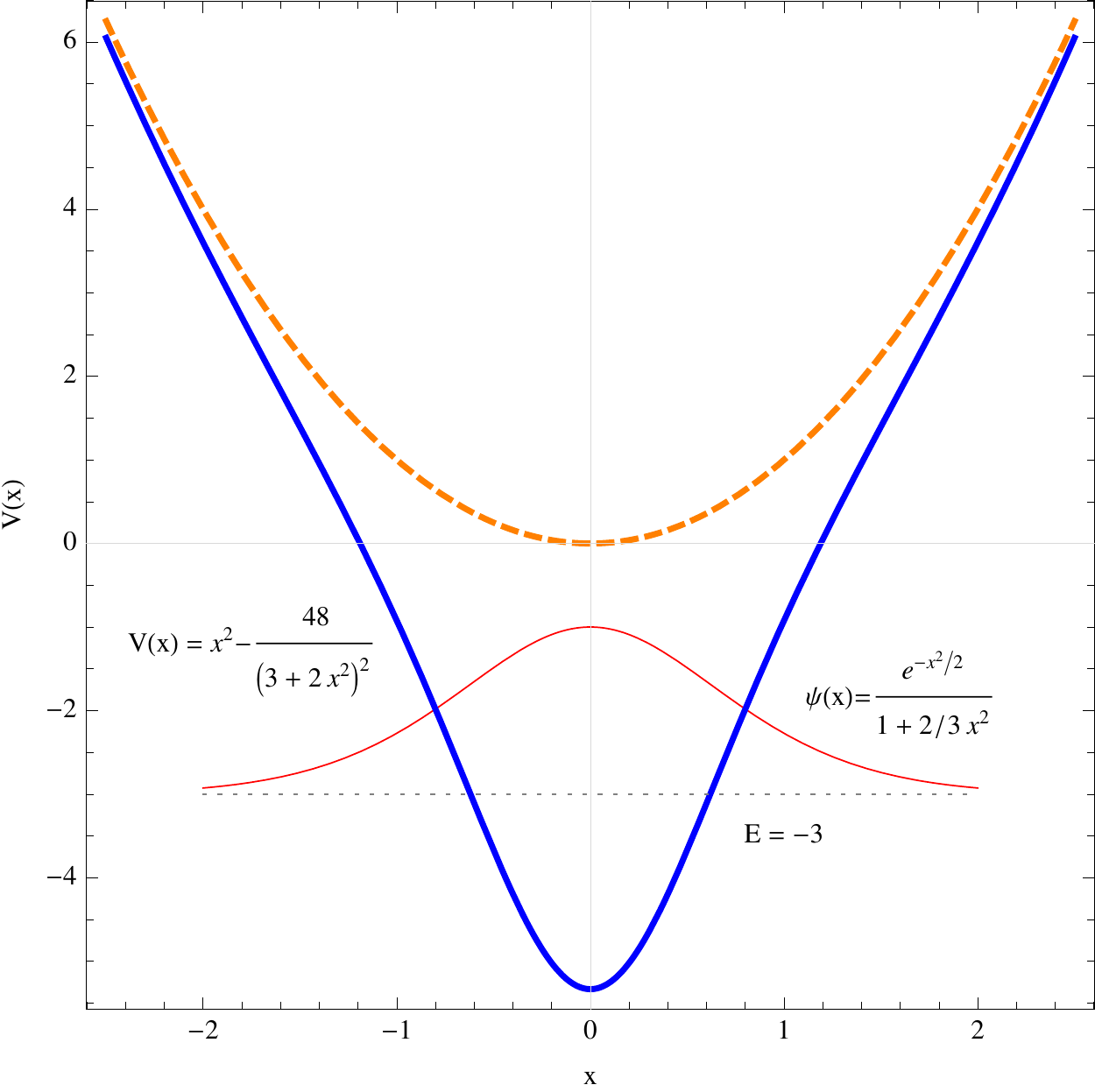}
\caption{\label{fig2}
The anharmonic potential (solid line) and the unperturbed harmonic potential energy curves and the
ground state wave function corresponding to ground state energy $E=-3$.}
\end{figure}

Let us now consider the rational anharmonic oscillator potential 
\begin{eqnarray}
V(x)  = x^2 - {{48}\over{(3+2x^2)^2}},
\end{eqnarray}
for which the ground state is exactly known
\[
\Psi_{gr}(x) \equiv  {{e^{- {{x^2}/2}}}\over {(1+{2\over 3} x^2)}}\;.
\]
This potential and the ground state associated with energy $E=-3$ are represented in Fig. \ref{fig2}, together with the
underlying harmonic oscillator potential (represented with dashed line). The advantage of knowing the ground state exactly is
that the claimed pointwise convergence of global-local method can be easily verified.

The corresponding moment equation for this system is
\begin{eqnarray}
\mu_{p+6} =& (E-3)\mu_{p+4} +\Big(p(p+7)+3E+{{39}\over 4}\Big) \mu_{p+2}  \nonumber \\
&+\Big(3p(p+3)+{9\over 4} E + 18\Big) \mu_p + {9\over 4} p(p-1)\mu_{p-2},
\end{eqnarray}
with $m_s = 5$. 
The corresponding recursion relation for Taylor's coefficients is
\begin{eqnarray}
\hspace*{-70.0pt} c_{n+2} = {{4(2n-E-7)c_{n-4}+4(n(11-n)-3E-15)c_{n-2}-3(13+3E+2n(3n-5))c_n}\over{9(n+1)(n+2)}}. \nonumber \\
\end{eqnarray}

\begin{table}
\caption{\label{table-rational}
Convergence for energy levels of the rational anharmonic potential
for various methods as a function of truncation order.
}
\centering
\begin{tabular}{ccccc}
 $N$ & $E_0$ & $E_1$ & $E_2$ & $E_3$ \\
 \hline
 \multicolumn{5}{c}{Direct OPPQ method with reference function $e^{-x^2/2}$}\\
 \hline
 20 & -2.919286247 & 0.910167637 & 3.603889662 & 5.913948003 \\
 40 & -2.996045597 & 0.799435213 & 3.437942536 & 5.720225364 \\
 60 & -2.999705662 & 0.792968365 & 3.426931961 & 5.705551770 \\
 80 & -2.999970901 & 0.792460681 & 3.426034090 & 5.704282111 \\
 100 & -2.999996463 & 0.792409589 & 3.425942480 & 5.704148274 \\
 \hline
 \multicolumn{5}{c}{Matching local and global behavior through  Eq. (18)}\\
 \hline
 20  & -3.192388811 &-0.534923547 & 3.546551239 & 5.743802129 \\
 40  & -3.008483327 & 0.735594478 & 3.407202508 & 5.604497284 \\
 60  & -3.000567710 & 0.789212601 & 3.424198982 & 5.696740280 \\
 80  & -3.000052234 & 0.792139749 & 3.425751715 & 5.703466230 \\
 100 & -3.000006028 & 0.792374246 & 3.425907672 & 5.704054764 \\
 \hline
 \multicolumn{5}{c}{Ground state wave function as reference function}\\
 \hline
 10 & -3.000000000 & 0.757672016 & 3.265699150 & 5.407722347 \\
 20 & -3.000000000 & 0.790563451 & 3.418884509 & 5.691130493 \\
 30 & -3.000000000 & 0.792220915 & 3.425272253 & 5.702879442 \\
 40 & -3.000000000 & 0.792377300 & 3.425841380 & 5.703957354 \\
 50 & -3.000000000 & 0.792398005 & 3.425914514 & 5.704099238 \\
 60 & -3.000000000 & 0.792401433 & 3.425926384 & 5.704122686 \\
\end{tabular}
\end{table}

Three representative calculations are given in Table \ref{table-rational}.
The first is OPPQ using the weight ${\cal R}(x) = e^{-{x^2/2}}$, the asymptotic form for the bound states. The second is 
obtained by implementing the global-local ansatz given in Eq.(18), ensuring not only accurate and fast converging 
energy eigenvalues, but also faithful representation of the wave function that has implicitly both the correct local and
global behavior. The third set of results illustrates the freedom of choice of the reference function within OPPQ method,
by taking it to be the exactly known ground state.

\begin{figure}[h]
\centering
\includegraphics[width=4.5in]{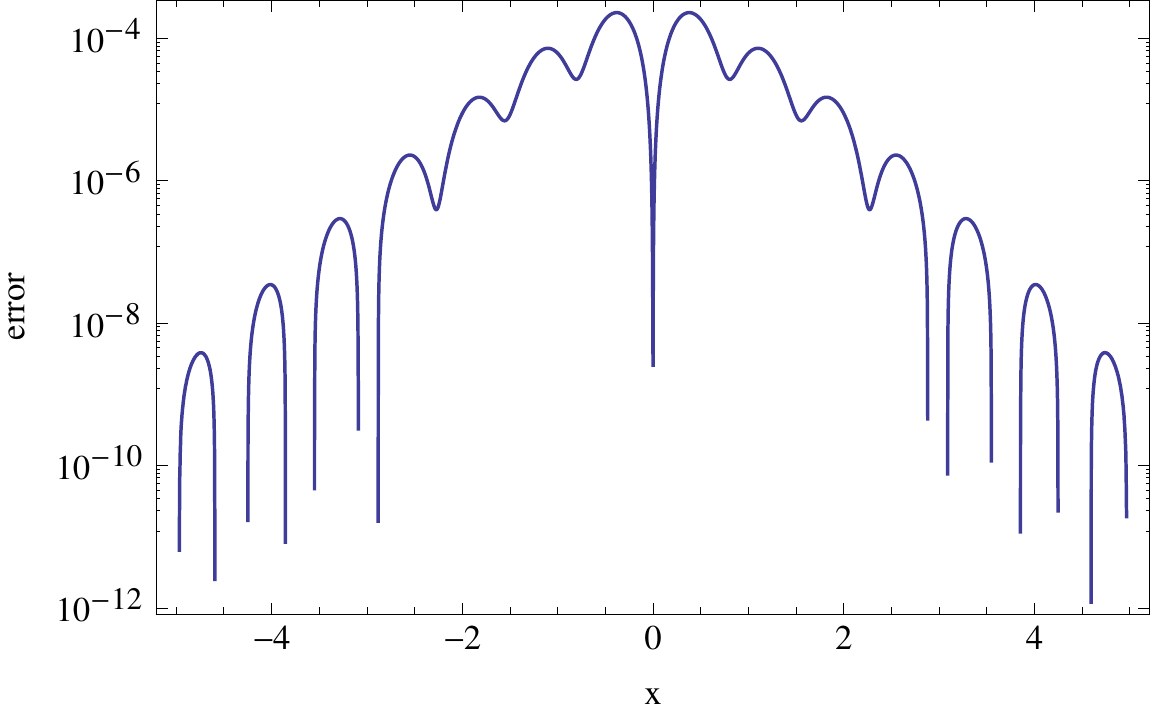}
\caption{\label{fig4}
The difference between the ground state wave function $\Psi_{\mbox{gr}}$ and its reconstruction obtained
by using Eqs. (9) and (18).}
\end{figure}

Figure \ref{fig4} demonstrate the pointwise convergence properties for the 
wave function calculated using the global-local quantization. The results presented are obtained 
by using Eqs. (9) and (18) with  truncation order $N=80$.
It is also instructive to compare the power expansion of the reconstructed
ground state wave function, scaled by the reference function ${\cal R}(x) = e^{-{x^2/2}}$, which is given by
\[
\Psi_{\mbox{rec}}(x) /e^{-\frac{x^2}2} = 1 - 0.662562 x^2 + 0.420992 x^4 - 0.237655 x^6 + {\cal O}(x^8)\quad,
\]
with the corresponding exact expansion of the ground state:
\[
\frac 1 {(1+{2\over 3} x^2)} = 1 - 0.666666 x^2 + 0.444444 x^4 - 0.296296 x^6 + {\cal O}(x^8) \quad.
\]
This shows that the solution provided by the global-local quantization has the claimed pointwise convergence properties
of the wave function together with fast convergence of energy levels.

\begin{figure}[h]
\includegraphics[width=4.5in]{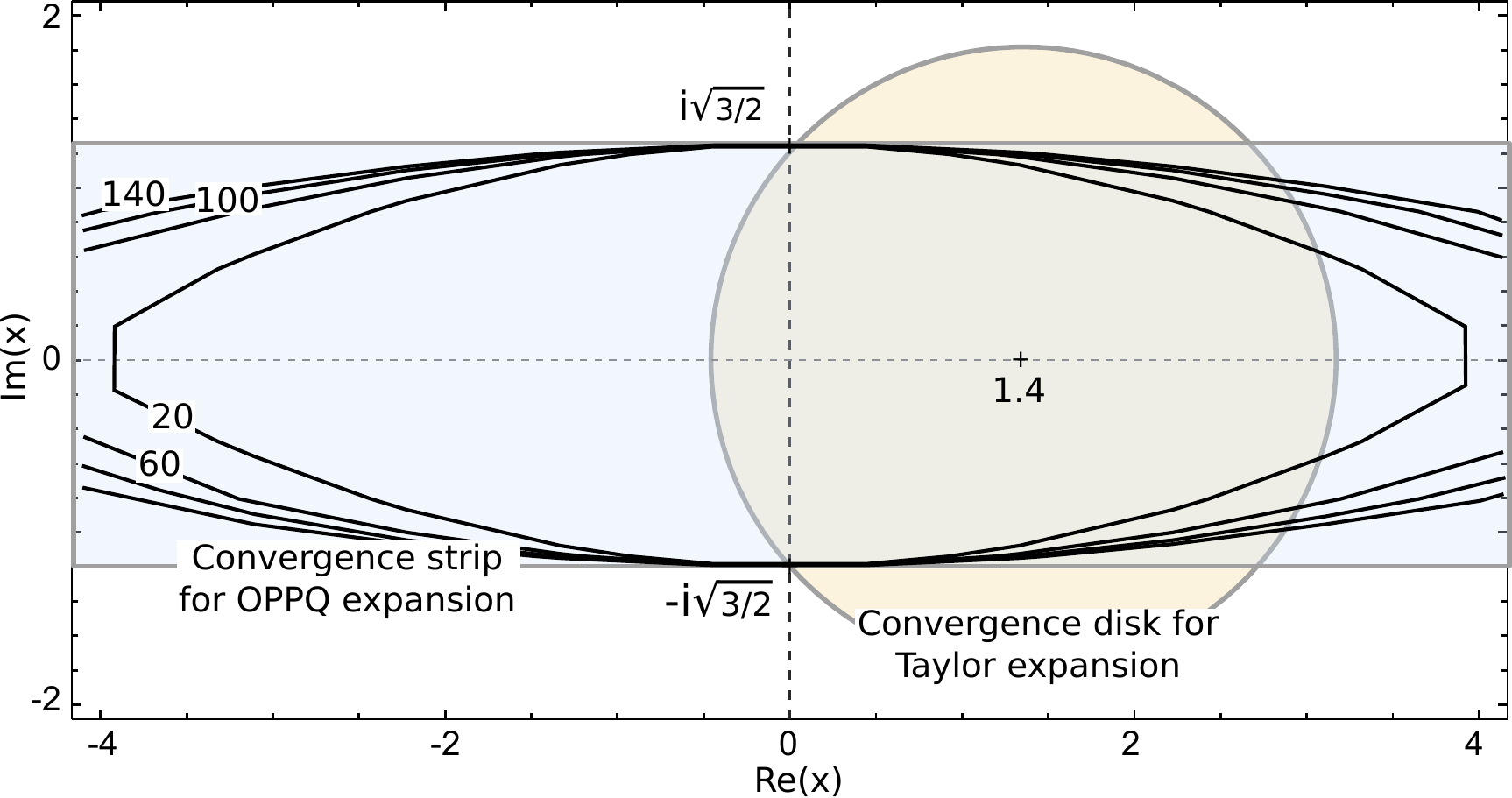}
\caption{\label{fig3}
Converge domains in the complex plane for Taylor's expansion and OPPQ expansion for the ground state wave function
scaled by the reference function ($\Psi_{gr}/{\cal R}$)
of the rational anharmonic oscillator in Eq. (19). The OPPQ expansion is carried out for the sequence of truncation
orders $N = 20, 60, 100$ and 140, showing that in the $N \rightarrow \infty$ limit,  its convergence domain is a horizontal
strip bordered by the singular point of the wave function. In contrast, Taylor's expansion does not
converge outside the disk centered at the chosen expansion center, in the calculated case $x_0 = 1.4$.
}
\end{figure}

It is instructive to compare the (non-uniform) global convergence (on the real line) of the OPPQ representation with the local Taylor series expansion. Specifically, the wave function representation in
terms of orthogonal polynomials as in Eq. (16) has global convergence properties, as opposed to the local convergence
of Taylor's power expansion which is always restricted to a disk. This point is illustrated in Fig. \ref{fig3} where the
convergence domain in the complex $x$-plane of $\Psi_{gr}$ covers an increasing horizontal strip, as the OPPQ truncation order in Eq. (16)
increases from $N=20$ to $N=140$. The strip is determined by the $\pm i\sqrt{3\over2}$ singularities. On the other hand, the
convergence of Taylor's expansion of the ground state is always local, limited by a disk around the expansion center
(chosen in Fig. 4 to be centered at 1.4) and bordered by the singularities.

The process of analytic continuation of a function, $f(x)$, results in power series expansions for the local expression
$f(\tau+\chi) = \sum_j c_j(\tau) \chi^j$, involving non-global expansion coefficients (i.e. the $c_j(\tau)$ are $\tau$ dependent). The orthonormal polynomial expansion in Eq.(19) transcends this into a global statement, particularly close to the real axis. Of course, one advantage of the conventional analytic continuation process is that one can control the uniformity of convergence of the analytic continuation to the target function. For our purposes, the non-uniform nature of the pointwise convergence of the OPPQ expansion is not a problem, since we are using the local information (i.e. the derivatives, etc.) at a chosen  point.

\section{General Considerations}

Traditional orthonormal basis expansion methods  in quantum mechanics, 
$\Psi(x)  = \sum_n  d_n {\cal B}_n(x)$ (i.e. $\langle {\cal B}_m|{\cal B}_n\rangle = \delta_{m,n}$) are not designed to recover the pointwize structure of the approximated solution, since they emphasize an $L^2$ convergence. That is, the $N$-th partial sum, $\Psi_N(x) = \sum_{n=0}^N d_n {\cal B}_n(x)$, converges to the physical solution according to  $\lim_{N\rightarrow \infty}\int dx\  |\Psi(x) - \Psi_N(x)|^2 = 0$. This does not imply pointwize convergence since in the infinite limit one could have $\Psi(x_j)-\Psi_\infty(x_j) \neq 0$  on a subset of measure zero.  The OPPQ representation has a greater chance of pointwize convergence  because its underlying structure mimics the usual power series expansion. In addition, it  can simultaneously recover the local, global, and asymptotic features of the desired physical solution.

The OPPQ basis functions $P_n(x){\cal R}(x)$ are non-orthonormal. The advantage of this representation is that the $\Omega_j$ projection coefficients are easily determined for the types of (multidimensional) systems of interest to us, although the present work is limited to one dimensional problems.

We note that $P_n(x)\sqrt{{\cal R}(x)}$ is expected to correspond to a complete orthonormal basis; however the expansion $\Psi(x) = \sum_n \gamma_n P_n(x)\sqrt{{\cal R}(x)}$ does not lead to an easy, algebraic (closed form) generation of the projection coefficients, $\gamma_n = \langle P_n(x)\sqrt{{\cal R}}|\Psi\rangle$, for arbitrary ${\cal R}$, as pursued here. Despite this, ${{\Psi}\over{\sqrt{\cal R}}} =  \sum_{n=0}^{\infty} \Omega_n P_n(x) \sqrt{{\cal R}(x)}$,  will correspond to a conventional orthonormal basis expansion, with expected $L^2$ convegence :

\begin{eqnarray}
Lim_{N\rightarrow \infty} \int dx\  |{{\Psi(x)}\over{\sqrt{{\cal R}(x)}}} - \sum_{n=0}^N\Omega_n P_n(x) \sqrt{{\cal R}(x)} | ^2 = 0.
\end{eqnarray}
Given that $Min_x{1\over{{\cal R}(x)}} > 0$ (a positive global minimum),  we then also have
\begin{eqnarray}
Lim_{N\rightarrow \infty} \int dx \ |\Psi(x) -\Psi_N(x)|^2 = 0,
\end{eqnarray}
where the OPPQ $N$-th partial sum is defined by
\begin{eqnarray}
\Psi_N(x) = \sum_{n=0}^N \Omega_n P_n(x) {\cal R}(x).
\end{eqnarray}
That is, the OPPQ representation will also be $L^2$ convergent.

The relation in Eq.(23), upon squaring and taking the corresponding integrals, gives the important result
\begin{eqnarray}
Lim_{N\rightarrow \infty} \sum_{n=0}^N \Omega_n^2 = \int dx \ {{\Psi^2}\over {\cal R}},
\end{eqnarray}
or
\begin{eqnarray}
Lim_{n\rightarrow \infty} \Omega_n = 0.
\end{eqnarray}
That is, if the basis $\{ P_n(x) \sqrt{{\cal R}(x)} \}$ is complete, then the limit in Eq. (26) holds.

An alternative representation for the orthonormality relations of the $P_n$'s is

\begin{eqnarray}
\int dx \ x^p P_n(x) {\cal R}(x) = 0,
\end{eqnarray} for $p < n$, leading to
\begin{eqnarray}
\mu_p = \int dx \ x^p \Psi_N(x), \  0 \leq p \leq N.
\end{eqnarray}
Thus the $N$-th OPPQ partial sum has its first $1+N$ moments identical to that of the physical state. This is the more general interpretation of the equality in Eq.(9).  Thus, the $\Psi_N(x)$ contain physical information.

Depending on the  asymptotic decay of the reference function, as compared to the complex extension of the Fourier kernel $e^{-ikx}$, the Fourier transform of the truncated OPPQ expression, ${\hat \Psi}_N(k)$,  can be an entire function, bounded along the real axis.  Furthermore, both the Fourier transform of the truncated OPPQ expression,${\hat \Psi}_N(k)$, and the actual Fourier transform of the physical solution, ${\hat \Psi}(k)$, will have identical power series expansions up to order $k^N$. It is therefore reasonable to expect that $|{\hat \Psi}(k) - {\hat \Psi}_N(k)| < \epsilon_N$ over some interval $|k| < \kappa_N$, where $Lim_{N\rightarrow \infty} \epsilon_N = 0$ and $Lim_{N\rightarrow \infty} \kappa_N = \infty$ . Therefore, the error in local approximation, near the origin in configuration space, is controlled by $|\Psi(x)-\Psi_N(x) | < {1\over{\sqrt{2\pi}}}|\int dk \ e^{ixk} ({\hat\Psi}(k)-{\hat\Psi}_N(k))| < {1\over{\sqrt{2\pi}}} \Big( \epsilon_N \kappa_N+ \int_{|k|> \kappa_N}dk \ |{\hat\Psi}(k)-{\hat\Psi}_N(k)|\Big)$. If $Lim_{N\rightarrow \infty}\epsilon_N\kappa_N = 0$, and ${\hat \Psi}_N(k)$ can capture the form of the physical solution's decay, for $|k| > \kappa_N$, then we can expect good local approximations for $x \approx 0$. In other words, how the reference function is chosen will lead to enhanced convergence rates to the local properties of the wavefunction in configuration space. More generally, it is to be expected that the OPPQ representation will generally converge, pointwise, to the physical solution.

\section{Conclusion}
The importance of moment representations in algebratizing many quantization problems is often overlooked. The importance, and flexibility, of weighted orthonormal polynomial representations has been amply demonstrated here and in previous works with regards to their ability to generate the discrete energies to arbitrary precision. The relevance of such representations for reconstructing the wavefunctions is strongly suggested by the present work, including our ability to quantize by imposing global-local constraints on the OPPQ, weighted polynomial, expansion.

\section*{Acknowledgments}

Discussions with  Dr. Daniel Bessis, Dr. Donald Kouri, and Dr. Walter Gautschi are greatly appreciated. Additional correspondences with Dr. H. N. Mhaskar and Dr. D. Lubinsky are also appreciated. One of the authors (DV) is grateful for the support received from the National Science Foundation
through a grant for the Center for Research on Complex Networks (HRD-1137732).

\section*{References}
1. Mhaskar H. N.  1996 {\it {Introduction to the Theory of Weighted Polynomial Approximation}} (Singapore: World Scientific Pub. Co. Inc.)\\
2.  Handy C R and Vrinceanu D  2013 {\em J. Phys. A: Math. Theor.} {\bf{46}} 135202 \\
3. Banerjee K 1979 {\em Proc. R. Soc. Lond. A} {\bf 368} 155 \\
4.  Hautot A 1986 {\em Phys. Rev. D} {\bf {33}} 437 \\
5. Tater M and Turbiner A V 1993 {\em J. Phys. A: Math. Gen.} {\bf 26} 697 \\
6. Handy C. R. and Murenzi R. 1998 {\em J. Phys. A: Math. Gen.} {\bf 31} 9897\\
7.  Handy C R and  Bessis D 1985 {\em Phys. Rev. Lett.} {\bf 55}, 931 \\
8.  Handy C R, Bessis D, Sigismondi G, and Morley T D 1988 {\em Phys. Rev. A} {\bf 37},4557 \\
9.  Handy C R, Bessis D, Sigismondi G, and Morley T D 1988 {\em Phys. Rev. Lett.} {\bf 60},253  \\
10.  Lasserre J-B  2009 {\it { Moments, Positive Polynomials and Their Applications}} (London: Imperial College Press ) \\
11.  Chvatal V 1983 {\it Linear Programming} ( W. H. Freeman and Co.) \\
\end{document}